\documentclass[aps,prl,groupedaddress,showpacs,twocolumn,floatfix]{revtex4-1}

\usepackage{amsmath}
\usepackage{amssymb,bm}
\usepackage{graphicx}
\usepackage{epstopdf}
\usepackage[colorlinks=true]{hyperref}
\usepackage{enumitem}
\usepackage{lipsum}
\usepackage{physics}
\usepackage{mathrsfs}
\usepackage{comment}
\usepackage{csquotes}

\renewcommand\({\begin{equation}}	% quick macro for equation with numbers
\renewcommand\){\end{equation}}

\renewcommand\[{\begin{eqnarray}}	% quick macro for equation with numbers
\renewcommand\]{\end{eqnarray}}

\begin{document}

\title{\textbf{Biasing topological charge injection in topological matter}}

\author{Mostafa Tanhayi Ahari}
\thanks{M. T. A. and S. Z. contributed equally to this work.}
\author{Shu Zhang}
\thanks{M. T. A. and S. Z. contributed equally to this work.}
\author{Ji Zou}
\author{Yaroslav Tserkovnyak}
\affiliation{Department of Physics and Astronomy, University of California, Los Angeles, California 90095, USA}

\begin{abstract}
% For mesoscopic systems involving large number of topological charges, 
% fluctuations are ubiquitous and important when operating at finite temperatures. In this work, 

We explore the interplay between topologies in the momentum and real spaces 
to formulate a thermodynamic description of nonequilibrium
injection of topological charges under external bias.
% in gapped topological phases of matter hosting topological defects in real space. 
We show that the edge modes engendered by the  momentum-space topology can play a functional role of connecting the external reservoirs to the bulk transport of topological charges in the real space.
% which leads to a control over the injection rate by exploiting thermal fluctuations. 
We illustrate our general results with two examples: the spin-torque injection of skyrmions in an electrically-biased integer 
quantum Hall system, and the vortex injection in a  topological $p\,+\,i\,p$ superconductor coupled to  
heat reservoirs. 
Based on the universal fractional entropy of the Majorana zero modes bound to the vortices, their controllable injection proposed in this work could provide a route for creating and manipulating Majorana fermions. 
\end{abstract}
\maketitle

{\it Introduction.}\textemdash Topology is ubiquitous. 
In condensed matter, topology provides a description for knotted structures of quasiparticle bands in the momentum space, 
as well as various real-space defect configurations in ordered media. 
Moreover, the two often coexist. 
For example, the interplay between Weyl electrons and magnetic vortices
% ~\cite{Balents2017} 
has attracted much attention 
in the recently discovered family of magnetic Weyl semimetals.
The new tool of magnetic topological quantum chemistry
% ~\cite{Bernevig2020} 
will certainly reveal more materials 
with coexisting band topology and magnetic order, where magnetic defects also naturally dwell.
Such systems may exhibit new transport phenomena with applications in electronics and spintronics.

In this Letter, we study the interplay between real- and momentum-space topologies 
in two-dimensional gapped topological phases of matter hosting point-like topological defects in the real space.  These defects carry quantized topological charges defined by the homotopy mapping of the order-parameter fields~\cite{Mermin1979}.
Obeying the topological conservation law, the flow of topological charges can be particularly robust, enabling long-distance transport useful for information transmission~\cite{Yaroslav}.
The bulk band topology dictates the existence of gapless modes localized at the edges, 
according to the bulk-boundary correspondence~\cite{Bernevig,Volb}.
We suggest a general functional aspect of topological media in addition to their bulk-edge transport.
The gapless edges, when biased by external sources, can serve as interconnects for controlling the bulk topological charges.
We illustrate this idea with two examples: an integer quantum Hall semiconductor and a two-dimensional topological $p\,+\,i\,p$ superconductor.

The bulk of the integer quantum Hall state at filling factor $\nu = 1$ is an insulating ferromagnet, if we neglect the Zeeman energy.
Its lowest-lying charged excitation is a skyrmion~\cite{Sondhi}, which is a real-space topological spin texture.
In response to an electrical bias, the chiral edge mode develops a ballistic current. 
This scenario is geometrically analogous to running an electric current through a metal in contact with a thin film of magnetic insulator, which results in skyrmion pumping into the magnet~\cite{Hector}.
In the quantum Hall system, we  reduce the Ohmic energy loss in generating the skyrmion current by exploiting the otherwise dissipationless edge mode.

In general, depending on the transport aspects of the edge modes, we can also utilize other means of biasing instead of an electric voltage.
Thermal bias can be generally applicable to different topological systems with heat conducting edges, which we explore with our second example.
The two-dimensional topological $p\,+\,i\,p$ superconductor supports chiral Majorana edge states. 
As a manifestion of the interplay between momentum- and real-space topology, 
a Majorana zero mode (MZM) arises as a bound state in a vortex excitation of the superconducting phase~\cite{Vol1}. 
We show in this case that the edge in contact with an external heat source can facilitate the injection of vortices and their associated Majorana modes into the bulk.
This could potentially provide a method for controllable creation and manipulation of Majorana fermions, for the purpose of quantum computing~\cite{Nayak}.

In both cases, the edge does work on the bulk dynamics and thus becomes dissipative.
This is usually undesirable for being detrimental to the quantized transport properties.
In systems with both momentum- and real-space topology, however, 
we point to the possibility of utilizing this interplay to create and manipulate topological charges, such as skyrmions and vortices, 
which are promising candidates proposed for future spintronic and quantum devices for information storage and processing~\cite{Friedman}.
We also discuss the thermodynamic description of the facilitated defect nucleation in general, as a nonequilibrium process.

{\it Stimulated nucleation of topological charges.}\textemdash We focus on the real-space topology in this section and consider the thermally activated nucleation of dilute topological charges at the boundaries, which thereafter enter the bulk as stable particle-like objects protected by topology. Let us write
the net topological charge flux injected into the bulk as
$J = \gamma - \rho \bar{\gamma}$, taking into account the nucleation rate $\gamma$, and the escape rate  $\bar{\gamma}$ per unit charge density $\rho$.
If in thermal equilibrium $J=0$, the ratio $\gamma/\bar{\gamma}$ is given by the equilibrium charge density,
\begin{equation}
     \rho_0  \propto e^{- \beta F_0},
     \label{eq:Boltzmann}
\end{equation}
which is controlled by the Boltzmann factor with
$F_0\gg k_BT$ being the free energy needed to create a unit of  topological charge
and $\beta = 1/k_B T$ at temperature $T$.

We now take the system out of equilibrium by coupling it to an external energy source.
If  a free energy kick could be provided accompanying the nucleation of the topological charges, the thermal excitation of them would become more likely.
For example, if a certain amount of work is done on the bulk order-parameter dynamics associated with the nucleation, a chemical potential $\mu$ can  be defined locally as the work done per unit charge, which shifts the energy exponent to
$e^{-\beta (F_0 - \mu)}$ in the Boltzmann factor~(\ref{eq:Boltzmann}). The ratio between the nucleation and escape rates is therefore enhanced by the fugacity:
\begin{equation}\label{genI}
    R \equiv \frac{\gamma}{\bar{\gamma}} =\rho_0 e^{\beta \mu},
\end{equation}
yielding a nonvanishing $J$ and a locally increasing $\rho$. A density gradient then develops across the bulk and drives the injected topological charges into diffusion. 
When they arrive at another unbiased edge, they are allowed to annihilate,  triggering an Onsager-reciprocal pumping process.
Therefore, instead of reaching an equilibrium with an  uniformly shifted charge density, the system eventually reaches an inhomogeneous steady state with a finite topological charge current $J$ in the bulk. In the linear-response regime, $J \propto \mu/k_B T$.

An alternative way to provide a thermodynamic kick is through a thermal bias, when there is a fixed entropy change $\Delta S$ associated with the injection of a topological charge. We assume the local temperature at the edge is $T_r$ in contact with an external thermal reservoir, higher than the reference (typically phononic) temperature $T$ in the bulk. To see the nucleation of topological charges is favored in this situation, we consider a charge passing into the bulk from the edge adiabatically. The amount of heat taken from the edge during this process should be $T_r \Delta S$, which  goes into the work $W$ on the topological charge. Therefore, we have 
% $\rho_0\rightarrow e^{-(E_0-W)/k_BT}\propto \rho_0e^{\delta T \Delta  S/k_BT}$, or equivalently
\( R=\rho_0 e^{\delta T\Delta S/k_BT}, \label{eq:R-S} \)
suggesting that the nucleation is favored if $\delta T\equiv T_r-T>0$, where the excitation gap  $F_0$ that controls $\rho_0$ remains unchanged. 
In the reaction theory, this way to bias the reaction rate is known as the Erying equation, which provides an approach to measure the entropy change~\cite{Erying}.
Similar to the discussion above, the stimulated injection results in a topological charge current in the bulk $J \propto (\Delta S/k_B) \delta T / T$ in the steady state.

In the two detailed examples to follow, we show how the edge modes engendered by the momentum-space topology can effectively induce the abovementioned chemical potential $\mu$ or the entropy change $\Delta S$ under a nonequilibrium bias. In this way, the edge modes play the role of connecting the external sources  to the bulk transport.

\begin{figure}
  \includegraphics[scale=.22]{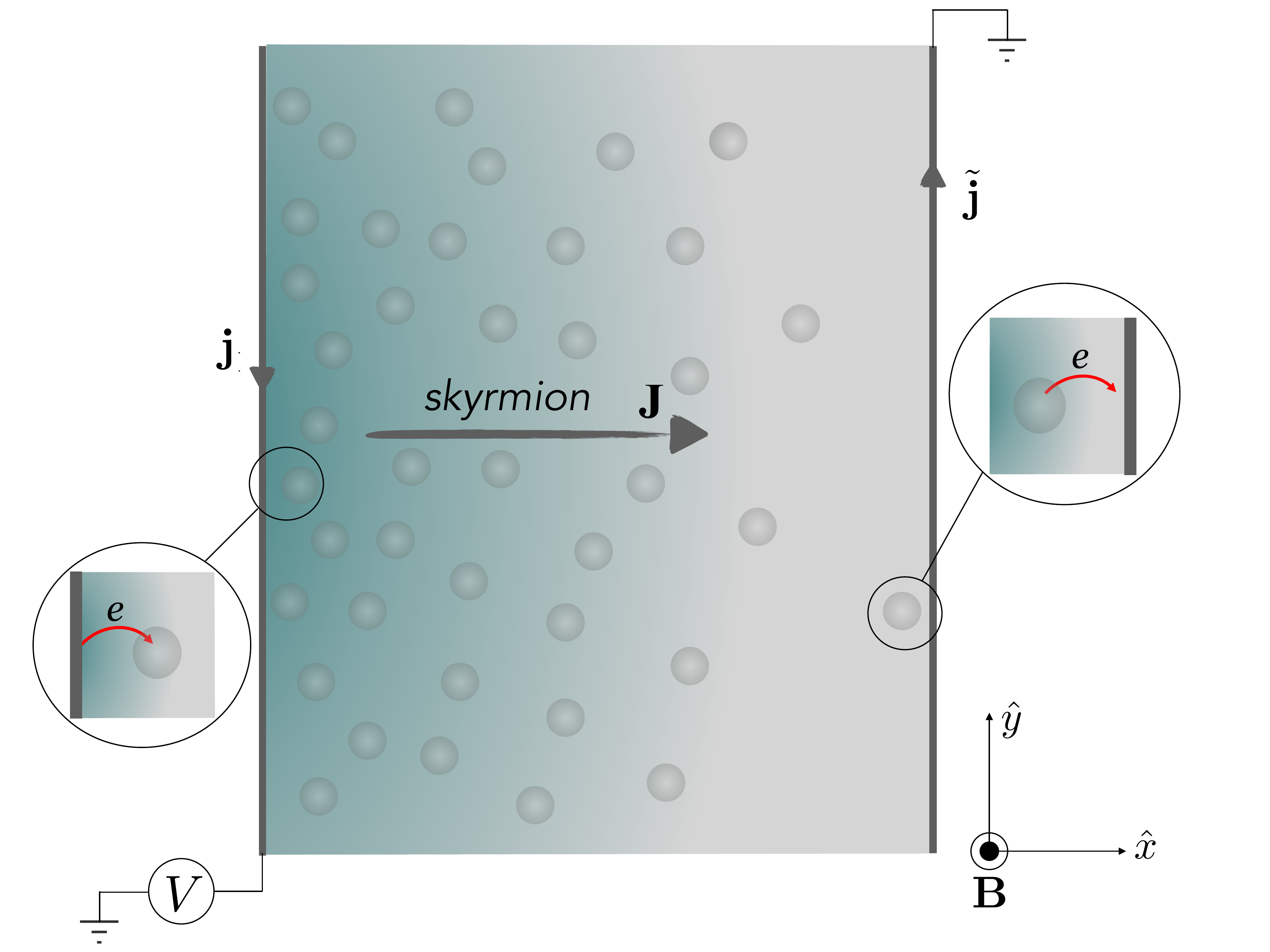}
   \caption{Skyrmion injection in quantum Hall system with a 2D strip geometry where the left edge is kept at a constant bias voltage $V$. Such a setup facilitates a transport of electric charge $e$ per skyrmion across the bulk, as shown in the insets. At the right edge, the current is induced by the outflow of skyrmions.}  
   \label{figqhe}
\end{figure}

{\it The quantum Hall system.}\textemdash We first test our idea with the $\nu = 1$ integer quantum Hall system by reproducing some of its well-known transport properties. To this end, we consider a quantum Hall strip in contact with a source electrode held at voltage $V$, and a grounded drain electrode.
% Let us consider an integer quantum Hall system with a filled lowest Landau level that is edge-coupled to a reservoir with voltage $V$ on the left, 
See Fig.~\ref{figqhe}. 
The left edge state flowing out of the source electrode carries the Hall current.
% At the boundary between the quantum Hall strip exists a gapless chiral edge state. 
% The gapless edge state is dissipationless---when biased by a voltage $V$ by connecting 
% to a lead at a particular point the entire edge equilibrates with the reservoir voltage which establishes a steady quantum Hall edge currents: 
When the entire edge is equilibrated with the voltage $V$
% V is negative
of the charge reservoir, the Hall conductance is exactly quantized to $\sigma_H = e^2/h$ and the current ${\bf j}= V \sigma_H \, {\hat{\bf y}}$ is dissipationless.

In the bulk, exchange interactions between the electron spins
lead to ferromagnetism with a vectorial order parameter $\mathbf{n}$, where skyrmion excitations are characterized by their topological charge,
\begin{align}
     q = \frac{1}{4\pi}\int d^2 \mathbf{r} \,\,  \mathbf{n} \cdot \left( \partial_x \mathbf{n} \times \partial_y\mathbf{n} \right).
\end{align}

We therefore consider the following dissipation channel.
% Next, we determine the effect of the controlled edge current on the bulk excitations of the quantum Hall ferromagnet. 
% On general grounds \cite{Ralph,Zhang} an 
The spin-polarized electric current $\mathbf{j}$  along the edge exerts a spin-transfer torque on the ferromagnetic order parameter~\cite{Ralph,Zhang},
\begin{align}
    \bm{\tau} = - \frac{h}{4\pi e} (\mathbf{j} \cdot \nabla) \mathbf{n}.
\end{align}
In the presence of spin dynamics, this torque  performs work~\cite{Hector},
\begin{equation}\label{work}
    W
    = \int d \ell  dt \, \dot{\mathbf{n}}\cdot  \bm{\tau} \times \mathbf{n} 
    = \frac{h}{e} \int d \ell  dt \, 
    \hat{\mathbf{z}} \cdot \left(  \mathbf{J} \times \mathbf{j} \right)
    =\frac{h}{e} j  \Delta q,
\end{equation}
where $\mathbf{J} = - (1/4\pi) \left[ \left(\hat{\mathbf{z}} \times \nabla \right) \mathbf{n} \times \dot{\mathbf{n}} \right] \cdot  \mathbf{n}$ is the skyrmion current density, and the integral is taken along the edge. 
% It can be checked that net topological charge of skyrmions $q$ injected into the bulk is given by
% \begin{align}
    %  q=\frac{e}{4\pi}\int dS\,\,  \mathbf{n} \cdot \left( \partial_x \mathbf{n} \times \partial_y\mathbf{n} \right)\,.
% \end{align}
% This is the topological invariant that underlies the stability of a skyrmionic configuration to smooth deformations 
% of the quantum Hall ferromagnetic order parameter. 
This defines a local chemical potential for the skyrmions 
$\mu = \delta W / \delta q = (h/e) j$. Substituting in 
$j = V (e^2/h)$, we obtain 
$\mu = eV>0$, which is precisely the local electrochemical potential provided by the charge reservoir. As expected, each skyrmion, as the lowest-lying charged excitation in the quantum Hall system, carries one electron charge~\cite{Sondhi}. 

% Invoking quantized Hall conductance, $e^2/h=j/V$, we get $W=qV$. 
% The work done on the system changes the free energy, $W=\Delta F$ see \cite{SM}, and effectively reduces the energy barrier to nucleate a shyrmion, $F_0-qV$. 
% Substituting this into Eq.~\eqref{genI}, the skyrmion nucleation rate reads
The work done by the edge current facilitates the skyrmion nucleation, as described by Eq.~\eqref{genI},
\begin{align}
      R=\rho_0 e^{\beta eV}. 
\end{align}
The injected skyrmion current  
in the bulk can be described by a combination of the density-driven diffusion and the transverse gyrotropic motion due to its nonzero topological charge~\cite{Ji,Hector}. 
% Just as an edge current exerts a torque on the order parameter of the quantum Hall feromagnet, 
At the opposite edge of the strip, according to the Onsager reciprocity, the spin dynamics associated with the skyrmion annihilation induces an electromotive field~\cite{Bender}, 
\begin{align}
    \bm{\mathcal{E}}= - \frac{h}{4 \pi e}  
    (\nabla \mathbf{n} \times \dot{\mathbf{n}}) \cdot \mathbf{n},
\end{align} 
which drives the current, 
\begin{equation}
    \widetilde{j} = \frac{e^2}{h} \int  d \bm{\ell} \cdot \bm{\mathcal{E}} 
    = e \dot{q}.
\end{equation}
Again recalling each topological charge corresponds to one electron charge $e$, the skyrmion transport infers a partially charge-conductive bulk.
% The edge is no longer equipotential and the edge current becomes dissipative.
This reproduces the standard picture for the integer quantum Hall leakage at finite temperatures. 
% The process of topological charge injection could also be of relevance to the break down of the quantization above a critical current~\cite{Nachtwei}.
The quantization breaks down as the injected skyrmion current is nonvanishing even in the macroscopic limit. The electric leakage (and the associated backflow) is governed by skyrmion diffusion~\cite{Hector}, whose impedance scales with the strip width.
From our perspective, using the edge mode as an interconnect, we can effectively tune the skyrmion charge current in the quantum Hall bulk by an external electric charge reservoir. More complex multiterminal circuits therefore can  potentially be utilized for skyrmion manipulations.

\begin{figure}
  \includegraphics[scale=.32]{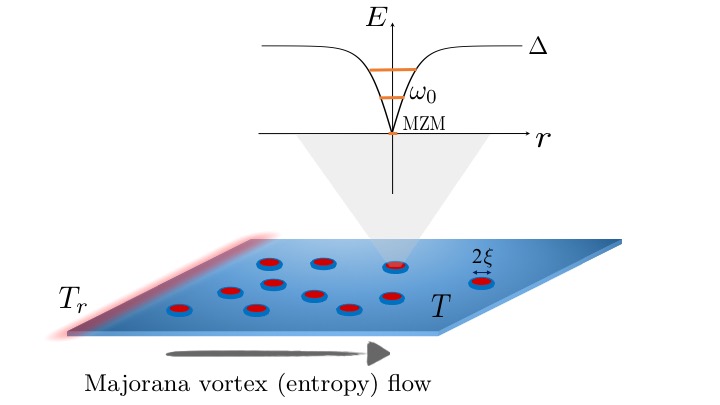}
   \caption{Schematic of a topological superconductor at a temperature $T$ whose edge is coupled to a heat reservoir at temperature $T_r=T+\delta T$. The thermal bias instigates an entropy flow carried by the MZMs, which are bound to the vortex core.
%   of size $\sim  \xi$.
    }  
   \label{figsc}
\end{figure}

{\it Topological $p\,+\,i\,p$ superconductor.}\textemdash As a second example, we consider vortex injection into a two-dimensional topological $p\,+\,i\,p$ superconductor, assuming spinless electrons and ordinary vortices with quantized flux $hc/2e$ (see Fig.~\ref{figsc}). Realization of such a topological phase has been proposed in various heterostructures~\cite{Ramon,Chiu,Sun,Sau}.
% Physically, the topological $p$-wave superconductivity is expected to emerge in 
% heterostructures realizing two-dimensional electron gas in which Cooper 
% pairing is established with only one active spin degree of freedom \cite{Ramon,Chiu,Sun,Sau}. 

Here, the edge states due to the momentum-space topology are gapless (for a thermodynamically large system) chiral Majorana states. While electric bias is no longer an practical option for these charge-neutral edge states, they are amenable to thermal bias. 
% To that end, we couple the edge of the superconductor 
% to a heat reservoir at temperature $T_r$, which is affecting the temperature of electronic
% degrees of freedom. 
As we will show in the following, the Majorana edge states can mediate an entropically favored injection of vortices.
% entropic effects will be responsible for the modified injection rate.
This is made possible by the presence of the Majorana bound states in the vortex cores. For a vortex far away from the edges, the energy spectrum of the  bound states is roughly given by 
$E_n \sim ( \Delta^2/E_F ) |n|$~\cite{Vol1,Roy}, where $\Delta$ is the superconducting pair potential in the bulk, $E_F$ is the Fermi energy, and $n$ takes integer values. There exists one MZM, separated from the first excited level by the minigap 
$\omega_0 =  \Delta^2/E_F $.

We set the system at a low temperature $k_B T \ll \omega_0 \ll \Delta$, such that the excitation of either Andreev bound states in vortices or Bogoliubov quasiparticles in the bulk is negligible. In this setting,  consider the adiabatic injection of a single vortex from the edge. 
After the injection, the MZM bound to the vortex core contributes an entropy increase $\Delta S = (1/2)k_B\ln{2}$ to the bulk, which is the universal fractional entropy of an MZM~\cite{Stone2}. 

Microscopically, the entropy increase in our scheme can be captured by a simple impurity model, where an MZM is coupled to the chiral Majorana edge mode.
Within the energy window of thermal fluctuations, we can safely treat the edge mode as linearly dispersing, with a constant density of state $\nu_0 \propto 1/\Delta$~\cite{Alicea}.
We consider the hybridization of the MZM and the edge mode via a local hopping interaction $t$,
which leads to the MZM density of states in the Breit-Wigner form:
\begin{equation}
    \nu (\omega) = \frac{1}{\pi} \frac{\Gamma}{\omega^2 + \Gamma^2},
\end{equation}
where the hybridization energy $\Gamma = \pi \nu_0 t^2$.
The entropy contribution can then be evaluated using $S_M = -\partial F_M/ \partial T$ with the free energy $F_M = -(1/\beta)\int_0^\infty d \omega \, \nu(\omega) \ln [1+\exp(-\beta \omega)]$. Note that the integral is taken only over positive energies. 
Since the hopping amplitude $t$ is dependent on the distance of the vortex from the edge $d$~\cite{Meng}, the injection process effectively tunes the hybridization energy $\Gamma$ from a large value at $d \lesssim \xi$ to infinitesimal at $d \gg \xi$, where $\xi$ is the coherence length of the superconductor.
In these two limits we have, respectively, $S_M(\beta \Gamma \gg 1) \rightarrow 0$ and $S_M(\beta \Gamma \ll 1) \rightarrow (1/2) k_B \ln 2$, which gives the entropy increase $\Delta S$.
We focus on the effects due to the Majorana states here, neglecting other generic contributions to the entropy, such as the positional configuration of the vortices.

We now turn on the thermal bias by coupling the edge to an external heat reservoir at a slightly higher temperature $T_r = T+\delta T$, shown in Fig.~\ref{figsc}, assuming the thermalization between the edge and the bulk due to electron-phonon coupling is relatively slow. Associated with nucleation of each vortex, an MZM at temperature $T_r$ is injected into the bulk, delivering an energy transfer $T_r \Delta S$.  
This heat transfer from the Majorana edge into vortex motion is closely analogous to the heat extraction by the adiabatic demagnetization cooling~\cite{MC}. 
%For an intuitive picture, we make the analogy to the paramagnetic cooling. We simply consider a spin-$1/2$ embedded in an environment at temperature $T_r$. We can couple the spin to a magnetic field $H$. By tuning $H$ from from infinitely large to zero, the spin undergoes an entropy increase of 
%$\Delta S_s = k_B \ln 2$ and draws the a total amount of heat $T_r \Delta S_s$ from the environment, resulting in the cooling effect. Here, $\Delta S_s$ can be calculated directly using the partition function of a two level system splitted by the Zeeman energy.
For this thermally biased topological superconductor, the tuning parameter that enables the energy transfer is the hybridization energy between the MZM and the edge mode during the vortex injection, as discussed above. We therefore emphasize again on the functionality of the edge mode as an interconnect between the heat reservoir and the bulk, which is essential in realizing the facilitated injection of topological charges.

The energy transfer $T_r \Delta S$, compared with that in the equilibrium $T \Delta S$, can effectively lower the energy barrier for
vortex injection. The raised ratio between the nucleation and escape rates is
\begin{equation}
    R = \rho_0 e^{\left[ \left( 1/2 \right) \ln 2 \right] \delta T/T} 
    = \rho_0 2^{\delta T/2T},
\end{equation}
as given by Eq.~(\ref{eq:R-S}).
Taking a double-log plot of the vortex injection rate versus $\delta T/T$, we obtain a linear function with the slope governed by the universal fractional entropy of an MZM. This feature could be measured in experiments to detect topological phases in superconductors. 
Our thermal biasing scheme via the edge states could also be explored to generate a controllable current of MZM for braiding purposes, etc.

{\it Discussion.}\textemdash Our perspective has focused on the interplay between momentum- and real-space topologies, leaving out of the picture various other dissipation channels for the edge states, and the rich dynamical effects of the topological charges themselves, which generally exist in systems with or without momentum-space topology. 
In the following, we examine the experimental relevance of our perspective. 

In order to study the Majorana vortex transport that is affected by the MZM core state, a superconductor with a 
large minigap is required to isolate MZMs from other midgap states.  Moreover, a superconductor with a short superconducting coherence length $\xi$ can potentially reduce vortex pinning by impurity and disorder in the bulk. At low temperatures, the electron interaction with dilute 
thermal phonons leads to the broadening of the Majorana bound state, which is exponential with $\Delta$~\cite{Loss}, i.e., $\sim \text{exp}[-\Delta/k_\text{B}T]$, and hence unimportant in the regime of interest, $T\ll T_c$. 
For example, the topological superconductor ($\text{Li}_{1-x}\text{Fe}_x$)OHFeSe~\cite{Liu} with 
a small coherence length $\xi\simeq 1.4$ nm, high $T_c$ ($42 \,$K), and a large superconducting gap $2\Delta\simeq 20$ meV can be 
robust against temperature fluctuations, 
which may be an attractive test bed
for our perspective. The small Fermi energy of the superconductor $E_F\simeq 50$-$60$ meV, 
leads to a relatively large minigap $\omega_0\simeq 1$ meV, which unambiguously separates MZM from other low-lying core 
states and impurity effects at $T=0.4 \,$K and magnetic field of $10\,$T ~\cite{Liu}. 

It is clear that in practice, one has to take into account nonuniversal properties such as 
contamination from impurity states and sample geometry. For example, the vortex-entry energy barrier in an ordinary 
type\textendash II superconductor can vary with sample size and geometry~\cite{Wang}, which are measured through their effects on the hysteresis of the magnetization curves~\cite{Bean,Zeldov}. This suggests a simple route for exploring macroscopic signatures of the underlying quantum statistics in driven collective vortex dynamics through nontrivial geometries. In particular, to have a more precise control over the nucleation of vortices, one can tailor the edge geometry and the applied magnetic field to harness the vortex entry barrier.

In addition, the vortex motion is often accompanied by the spectral flow of the fermionic states bound to its core~\cite{Kopnin}, which could affect the thermal transport~\cite{Freimuth}. 
Here, we consider weak bias at the boundary and thus slow (quasistatic) vortex dynamics. In this limit, dynamical effects such as the spectral flow of states above the minigap in the vortex core should not significantly change our conclusions~\cite{Stonesf}. 
With the excitation of antivortices suppressed by an external magnetic field, only one species of vortices is assumed. We have also neglected the possible interactions between vortices, which may be justified since the Majorana tunneling is exponentially suppressed as a function the distance between vortices~\cite{Meng}. Depending on the global geometry, we may also need to keep track of the  winding superflow built up due to vortex flow and the backaction by the associated free energy~\cite{Dalton}. 

The advantage of considering the topological aspects is that the quantities of interest are ``quantized'', such as the chemical potential of skyrmions in the quantum Hall example, and the entropy per MZM in the topological superconductor example.
A half quantized entropy change was recently proposed to be measurable as the fingerprint signature of an MZM by coupling it to a metallic lead~\cite{Oreg}.
Here, having these definite quantities could enable us to effectively control the biased injection of topological charges. Furthermore, the nucleation timescale of the vortices that bind an MZM 
is exponentially affected by 
entropic effects, $1/R=(1/\rho_0)2^{-\delta T/2T}$. This serves as an example of 
enthalpy-entropy compensation, or the 
Meyer-Neldel rule \cite{Yelon,Wild} for topological charge nucleation. 

Since energy transfer, either in work or heat, is affected by noise in the leads and the order parameter variations, it may fluctuate between different realizations of the nucleation process. Fortunately, taking such fluctuations into consideration, our formalism still applies thanks to the well-established Jarzynski equality~\cite{Jar}. Instead of defining a chemical potential through the averaged work $\langle W \rangle$, we can think of the average of the exponentiated work $\langle e^{-\beta W} \rangle$ = $e^{-\beta \Delta F}$. The equality holds for nonequilibrium states, as well as in the presence of strong system-environment coupling~\cite{Jar2}. The biasing effects of $\Delta F$ can then be discussed similarly to what we have shown.

The examples we have discussed are based on the well-ordered phase of the bulk. Strong thermal fluctuations, on the other hand, can sometimes be more favorable for the injection of topological charges~\cite{Hedgehog}, with easily activated dynamics of the order parameters. For an understanding near the critical temperature, a dynamic critical theory is needed with stochastic terms and nonequilibrium drive~\cite{Hohenberg}, which is interesting for future studies. We remark that $\nu=1$ integer quantum Hall bilayer is another experimental platform for skyrmionic pseudospin textures, as well as vortex injection (due to the easy-plane anisotropy)~\cite{deannphy}.

In this work, we have presented a framework to trigger and control bulk transport of topological charge by thermodynamic biasing at the edges, which is governed by the universal properties of the topological charges. Along with the discovery of more topological materials, the interplay between the band topology and the real-space order-parameter textures takes place in a diverse range of settings. The edge states, which are relatively easy to access in experiments, provide a promising route to bias and control the transport of topological charges in the bulk.

\begin{acknowledgements}
It is a pleasure to acknowledge discussions with Hector Ochoa and Pascal Simon. This work was supported by the U.S. Department of Energy, Office of Basic Energy Sciences under Grant No. DE-SC0012190.
\end{acknowledgements}

\end{document}